\documentclass[12pt]{article}
\usepackage[font=small,format=plain,labelfont=bf,up]{caption}
\usepackage{epsfig, graphicx}
\usepackage{float}
\usepackage{amsmath}
\usepackage{latexsym}
\usepackage{epsfig, graphicx} 
\usepackage{graphics}
\usepackage{epsfig, graphicx}
\usepackage{amsmath}
\usepackage{latexsym}
\usepackage{amstext}
\newcommand{\mb}[1]{ \mbox{\boldmath$#1$} }
\newcommand{\ds}{\displaystyle}
\newcommand{\beq}{\begin{eqnarray}}
\newcommand{\eeq}{\end{eqnarray}}
\newcommand{\beqq}{\begin{eqnarray*}}
\newcommand{\eeqq}{\end{eqnarray*}}
\newcommand{\p}{\partial}

\newcommand{\x}{\mbox{\boldmath$x$}}
\newcommand{\w}{\mbox{\boldmath$w$}}
\newcommand{\n}{\mbox{\boldmath$n$}}

\newcommand{\y}{\mbox{\boldmath$y$}}
\newcommand{\z}{\mbox{\boldmath$z$}}

\begin{document}

\title{Modeling the early steps of cytoplasmic trafficking in viral infection and gene delivery}
\author{C. Amoruso, T. Lagache D. Holcman,\\ Department of Applied Mathematics and Computational Biology, IBENS
 \\Ecole Normale Sup\'erieure, 46 rue d'Ulm 75005 Paris, France}
\maketitle
\begin{abstract}
Gene delivery of nucleic acid to the cell nucleus is a fundamental step in gene therapy. In this review of modeling drug and gene delivery, we focus on the particular stage of plasmid DNA or virus cytoplasmic trafficking. A challenging problem is to quantify the success of this limiting stage. We present some models and simulations of plasmid trafficking and of the limiting phase of DNA-polycation escape from an endosome and discuss virus cytoplasmic trafficking. The models can be used to assess the success of viral escape from endosomes, to quantify the early step of viral-cell infection, and to propose new simulation tools for designing new hybrid-viruses as synthetic vectors.
\end{abstract}
\section{Introduction}
Drug delivery is a multi-step process of delivering a cocktail of molecules through different tissue levels to a specific location in the body, which is usually a difficult task. For example, the Brain Blood Barrier is a physical barrier that drastically limits the access of molecules to the central nervous system, preventing efficient drug delivery. The possibility of injecting to a cell a piece of DNA that can reach the nucleus, in order to synthesize a given protein, is a very seducing idea, which has driven gene therapy and genetic crop modification. Gene therapy has been tried in the quest to cure diseases caused by single-gene defects, such as cancer and hereditary diseases linked to a genetic defect \cite{Wikipedia}.

Non-viral methods for gene delivery use a variety of tools such as polymeric gene carriers, microinjection, gene gun, hydrostatic pressure, electroporation, continuous infusion, and many others
\cite{Wikipedia}. Gene delivery can be used, in principle, for fighting major diseases, ranging from viral infection to epilepsy. Although viruses are great predators, they can serve as vectors of drugs. For example, the Adeno-Associated Virus of type 2, (parvoviruses \cite{principleofvirology}) can carry gene coding for the neuropeptide Y, which is transduced in some areas of the hippocampus, where it is injected. The peptide is then released from neurons during high brain stimulations and reduces glutamate release, thus preventing the propagation of a high firing rate in neighboring neurons \cite{Kokaia}. This therapy has the potential to reduce epileptic form activity.

In this review, we discuss recent models of intracellular delivery and viral trafficking. Mathematical and physical models are constructed for the purpose of predicting and quantifying infectivity and the success of gene delivery. The models give rise to rational Brownian dynamics simulations for the study of sensitivity to parameters and, eventually, for testing the increase or the drop in infectivity, by using simultaneously a combination of various drugs. The modeling approach can be used for the optimization of the delivery in a high-dimensional parameter space.
In the first part of the review, we present several approaches to the study of some limiting steps of gene delivery. A major limiting step remains the success of plasmid escape from an endosomal compartment. In the second part, we present models of viral entry, the endosomal step and trafficking in the cytoplasm. These models can be used to predict the efficacy of new synthetic vectors, based on hybrid viruses.

\section{Mechanisms of gene delivery. }
\subsubsection*{Introduction: Two types of gene vectors}

\begin{figure} [htb]
\includegraphics[angle=0,width=1\textwidth]{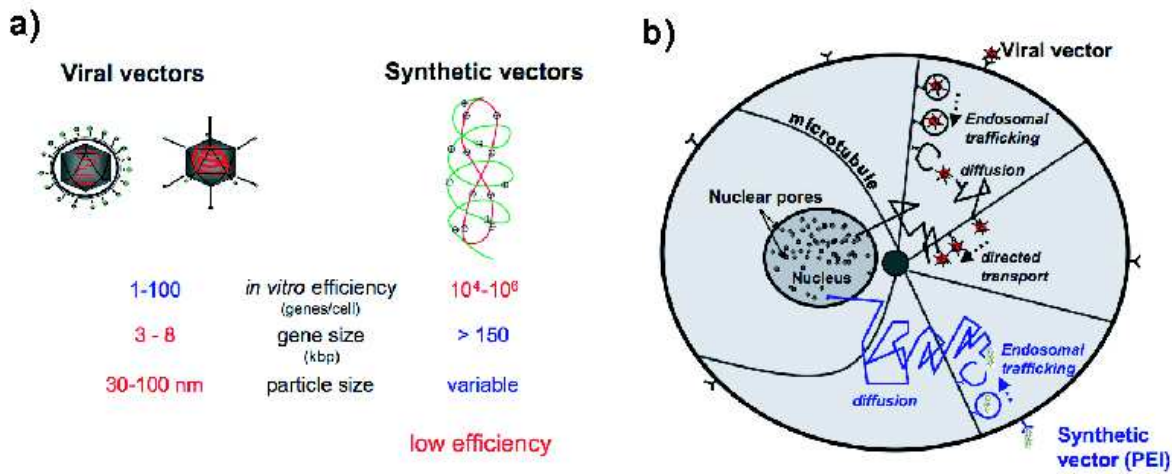}
\caption{\textbf{Viral and synthetic vectors are used to transfer genes in the nucleus}. {\bf (a)} characteristics of viral and synthetic vectors. While synthetic vector polycations are preferred to viruses, they are safer and the size of the genes is not limited by any capsid architecture, but their efficiency to escape endosomes and traffic through the risky cytoplasm is still very low. {\bf (b)} schematic description of early step of infection for viral and synthetic vectors. Synthetic vectors are not assisted by active transport during their cytoplasmic trafficking.}
\label{vectors}
\end{figure}

Gene therapy is a new therapeutic strategy that offers the promise
of treating diseases through insertion, alteration, or removal of
genes within an individual's cells and biological tissues. In
general, any drug molecule must reach its intended site of action to
exert its effect and to avoid non specific potentially dangerous interactions. A daunting hurdle of drug delivery is the
large amount of charged molecules such as DNA that need to reach the
cell nucleus. Indeed, the DNA molecule must initially pass through the
cell membrane, traffic inside the cell cytoplasm to finally enter
the  nuclear pore in a form suitable for transcription to start. To overcome efficiently these barriers, the genetic material is usually associated to a viral or synthetic vector (see figure \ref{vectors}).
Viruses have developed evolutionary tools to enter cells  and get
transported towards the different compartments such as the nucleus
and finally to reproduce using the protein synthesis machinery of
the host cell (see figure \ref{vectors}-b). However their use
presents several limitations: viral vectors are not safe and have
been implicated in death during clinical trials \cite{pb1}, they can
trigger an immune response and thus cannot be administrated
repeatedly. Finally, the size of the transferred genome is limited
by the nucleocapsid architecture. To overcome the difficulties
imposed by the use of viral particles as vectors, synthetic vectors
such as lipids, cationic polymers, peptides or combinations thereof,
have been developed in parallel. In particular, the most common used
vectors for nonviral gene delivery are cationic lipids, first
introduced in 1987 \cite{Felgner} and later on cationic polymers
\cite{WU}, followed by improved \cite{Behr} polycations such as the polyethylenimine
(PEI) for transfection experiments.  Cationic lipids are macromolecules consisting of an
aqueous core enclosed in a spherical phospholipid layer with
positively charged head groups, while polycations are positively
charged polymers which can have different structures (linear or
branched). Both molecules are able to condense and neutralize DNA
molecules through electrostatic interactions between their positive
charges and the negatively charged DNA phosphate groups. In both
cases, polycations protect DNA from undesirable degradation and
facilitate entry into cells.

Despite the high degree of reproducibility, the ease of modification, the lack of  toxicity and the ability to deliver large pieces of DNA, polycations and liposomes are still much less efficient than their viral
counterpart due to low cellular uptake and endosomal escape.  A new promising strategy  is the use of cationic liposomes incorporating fusogenic peptides from  viral glycoproteins, which are supposed to enhance  endosomal escape by mimicking  the mechanism of fusion of viral envelopes with host cell endosomal membrane \cite{vector1,vector2}.

Unraveling the molecular mechanisms that underly the cellular
behavior of both viruses and synthetic vectors is now needed to
develop and optimize efficient hybrid vectors. In particular, the endosomal escape remains a major barrier in gene delivery and prompt release from an endosomal compartment presumably constitutes one of
the critical steps in determining the efficiency of transfection. In
addition, the cytosolic motion of large DNA molecules is limited by
physical and chemical barriers of the crowded cytoplasm
\cite{Verkman,Dauty}: whereas molecules smaller than 500 kDa can
diffuse, larger cargos such as viruses or synthetic gene vectors,
require an active transport system \cite{sodeik} such as the
microtubules (MTs). Consequently, to understand how molecular
components of gene vectors can affect quantitatively their ability
to pass through these two main limiting barriers of gene expression,
we will present a review on our recent biophysical models for both viral and
synthetic vectors. Indeed, while modeling the endosomal escape of
viruses will help to understand the molecular mechanisms underlying
their reliable escape at a given pH, a model of the free cytoplasmic
step will allow to analyze the active transport along MTs, which increases the probability that a viral particle
reaches a nuclear pore in comparison with a Brownian synthetic gene vector.

\subsubsection*{The cellular uptake pathways}
Since the phospholipid bilayer of the plasma membrane has a
hydrophobic exterior and a hydrophobic interior, any polar
molecules, including DNAs and proteins, are unable to freely pass
through cell membranes. To circumvent this barrier,  an approach
consists in transferring naked nucleic acids by inducing a membrane
destabilization. An externally applied electrical field can
temporarily disrupt areas of the membrane  allowing polar molecules
to pass, then the membrane can reseal quickly and leave the cell
intact \cite{Electrop1,Electrop2,Electrop3}. This
method, commonly known as electropermeabilization or electroporation, provides versatility and efficiency,
but it suffers from non specific transport across the membrane which
can lead to ion imbalance and thus improper cell functions. It can
also lead to irreversible cell damage if the electrical field is too
strong. { Microinjection also permits rapid delivery of genes to the
cytosol or the nucleus by using a glass micropipette to inject
genetic material into cells.}

Permeabilization is another used technique of
delivery in which pore-forming agents which have the ability to fuse
with the membrane are used to form large apertures in the cell
membrane. However these techniques are highly invasive and cannot be used for
in vivo gene delivery. A different strategy to overcome the initial
cell membrane barrier is by associating genetic molecules with viral
or synthetic vectors such as lipids, cationic polymers, peptides or
combinations thereof, which are internalized through the endocytic
pathway. Endocytosis is the process by which cells absorb molecules
by engulfing and enclosing them into vesicles. There are several
different endocytic pathways:

\begin{enumerate}
\item {\it Clathrin-Mediated Endocytosis}
Clathrin-Mediated Endocytosis (CME) is a process by which cells
internalize molecules through the strong binding of a ligand to a
specific cell surface receptor. This process  results in the clustering of
the ligand-complexes in coated pits which are formed by cytosolic
proteins, the main unit being clathrin. The coated pits then
invaginate the plasma membrane to form clathrin-coated vesicles.

\item {\it Caveolae-Mediated Endocytosis} Caveolae are small
(approx. 50 nm in diameter) pits in the membrane that resemble the
shape of a cave which can mediate uptake of extra-cellular molecules
through complex signaling.

\item {\it Macropinocytosis}  A mechanism of endocytosis in which
large droplets of fluid are trapped underneath extensions (ruffles)
of the cell surface.

\item {\it Phagocytosis} is the process by which cells
bind and internalize large ($ >0.5  \mu$) pathogens such as
bacteria or micro-organisms.
\end{enumerate}

\subsubsection*{Endosomal escape}

{Once a gene vector (synthetic or viral) enters an endosome, it has to escape into the cytoplasm before being degraded  in lysosomes. Although the exact pathways leading to endosomal escape are not fully elucidated, they are limiting steps in gene delivery. Most viruses possess efficient endosomolytic proteins allowing them to disrupt the endosomal membrane, such as the VP1 penetration protein of the adeno-associated virus (AAV)  \cite{parvo} or the influenza hemagglutinin (HA), \cite{HA}. In addition, the biophysical mechanism leading to endosomal membrane destabilization and concomitant plasmids release for synthetic vectors is still poorly understood. However, in both cases, acidification
of the endosome is needed to trigger endosomal escape. For viruses protons or low pH activated proteases bind viral  endosomolytic proteins, triggering their conformational change into a fusogenic state \cite{HA,parvo}.
Polyplexes (DNA/cationic polymer) use a different strategy to escape endosomal compartments, based on
the so-called proton sponge mechanism. Indeed some polycations have a high buffering capacity in a wide range of pH
thus preventing the DNA from degradation and inducing an influx of chloride $Cl^-$ ions
to maintain electrical neutrality. The osmotic pressure due to the influx of  $Cl^-$ ions causes endosome swelling
and eventually membrane disruption. }

\subsubsection*{ Cytoplasmic trafficking}
Following the endosomal escape, viral and synthetic gene vectors
have to travel through the crowded cytoplasm to reach the nucleus
and deliver their genetic material through the nuclear pores. While
the cytoplasmic movement of viral particles towards the nucleus is
facilitated by the microtubular network and viral proteins, very
little is known about the fate of non-viral DNA vectors in  the
cytoplasm. However, trapping of large DNA particles ($>$500 kDa) in
the crowded cytoplasm drastically hinders their cytoplasmic
diffusion \cite{Verkman,Dauty} and subsequently diminishes the
transfection rate of synthetic gene vectors.
{In addition, one of the critical barriers in polyplex-mediated gene delivery is the timely unpacking process of the complexes within the target cell to liberate the DNA for efficient gene transfer \cite{DNA-unpack}. A desirable property of a gene carrier is that it strongly binds to DNA to form  more compact structures that can provide better protection of genetic material against nucleases and efficient transportation through the cytoplasm. However stronger polycation-DNA interactions produce a counter-productive effect on the timely release of DNA for transcription. Therefore an ideal vector would be capable of binding strongly to the DNA during the early stages of intracellular transport and to release the DNA right before nuclear entry.}

\subsubsection*{ Nuclear delivery}
Expression of therapeutic proteins in gene therapy requires that the transfection agent, a virus or a synthetic vector, delivers its cargo through the nuclear envelope (NE) into the cell nucleus. Later on, the genetic material will be involved in the cellular transcriptional machinery. {Some viruses such as AAV are able to transiently disrupt the NE \cite{AAV-nucleus}, delivering their intact capsid for disassembly inside the nucleus. However, most viruses translocate inside the nucleus through nuclear pore complexes (NPCs) by using nuclear localization sequences (NLSs). This facilitated nuclear import can accommodate the transport of molecules with diameters of up to 39 nm \cite{Paine,Peters}. Thus, small viruses such as hepatitis B virus \cite{HBV-nucleus} or baculoviruses \cite{bacu-nucleus} can cross the NPC and disassemble inside the nucleus. In contrast, larger viruses  such as HIV \cite{HIV-nucleus} or influenza \cite{Influenza-nucleus} uncoat inside the cytoplasm, or near a NPC, for the adenovirus \cite{adeno-nucleus}.
The released components contain NLSs and are thereby able to cross the NPC. Viral derived NLS peptides associated to polycations have also appeared to enhance the nuclear delivery of genes \cite{synt-nucleus}.

Recent biophysical models have been proposed to quantify the molecular mechanisms of viral facilitated transport through the NPC \cite{Lowe,Stewart,peters,wente}, however the exact import mechanisms still remain unclear. Interestingly, it has been found that pressure due to DNA condensation in double-stranded DNA viruses, such as the herpes virus, is as high as 50 atmospheres, allowing a direct injection of the genome inside the nucleus through the NPC \cite{gelbart}.

\section{Modeling the endosomal pathway in Gene Delivery with cationic lipids and polycations.}
The relative inefficiency of transfection using polycations,
compared to the use of viral vectors remains the largest barrier to
synthetic vector development and applications. Elucidating how
non-viral vectors behave at the intracellular level is therefore a
necessary step for synthetic vector improvement and optimization
\cite{synth}. Biophysical modeling of polyelectolytes in confined
domains, their interaction with biological membranes and escape from
endosomes would allow us to design new efficient vectors.
Computational models have accounted for endosomal escape using
kinetic equations \cite{Varga}, where parameters describing
multiples steps of gene delivery pathways are specified from
experiments, numerical fits or through  statistical
mechanics analysis \cite{ChemPhys}. However, it is difficult to extend these methods to identify the biophysical mechanisms underlying DNA endosomal escape.

\subsubsection*{Two biophysical scenarios to study the endosomal release of DNA}
Cationic polymers, such as PEI, condense DNA into nano-sized
polymer/DNA complexes (polyplexes) by a self-assembling process due
to electrostatic interactions of the positively charged polymer with
the negatively charged DNA. Polyplexes with positive surface charges
are formed when the number of positive charges of the polymer
exceed that of the negative DNA charges. When a polyplex encounters
the cell surface, it interacts  with the negatively-charged cellular
membrane and can be taken up into the cells via endocytosis. In the
intracellular environment the polyplexes are located in endosomes
that become acidified. In this case, DNA is prone to degradation by
lysosomal enzymes.
\begin{figure} [htb]
\includegraphics[angle=0,width=0.4\textwidth]{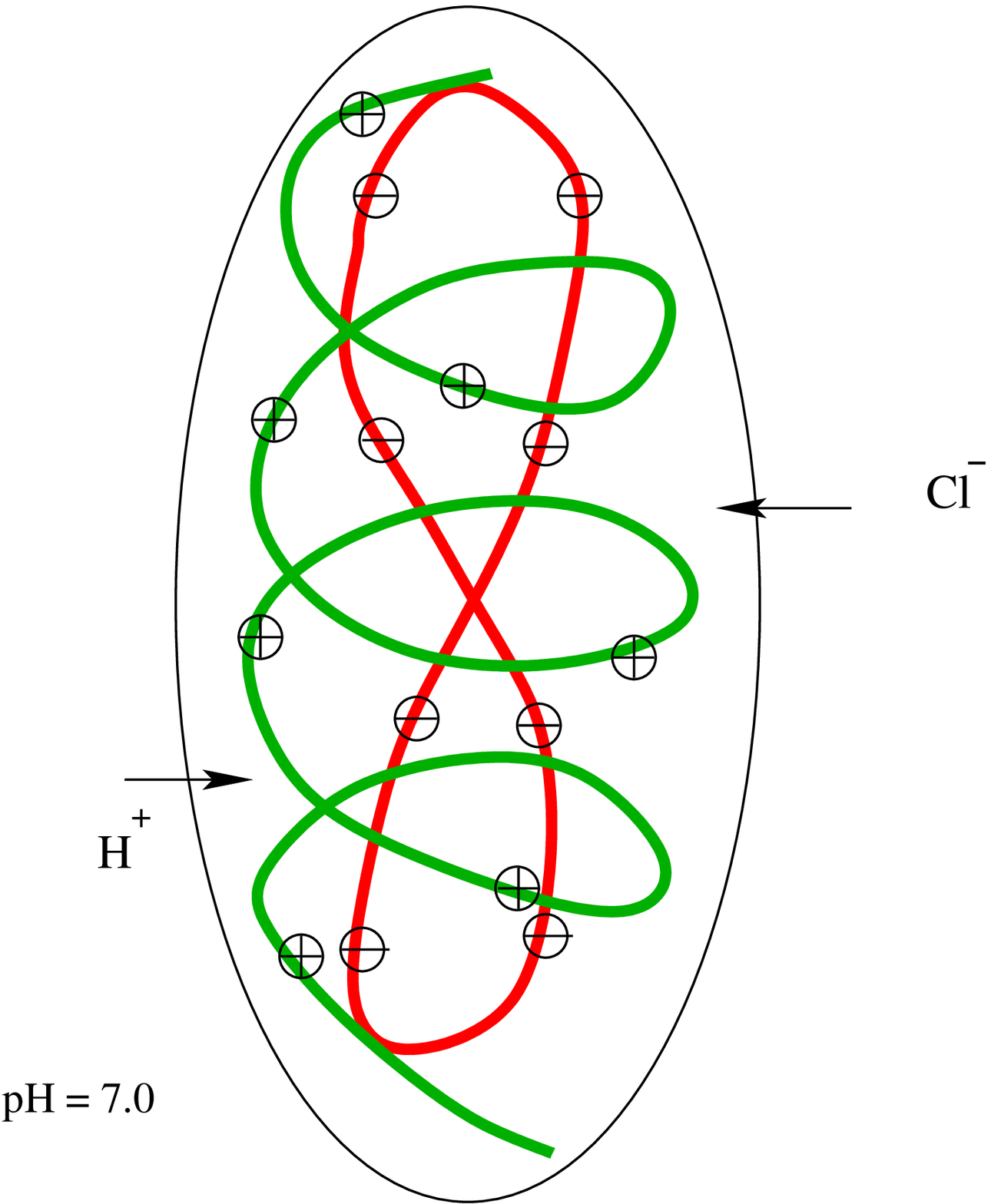}
\includegraphics[angle=0,width=0.4\textwidth]{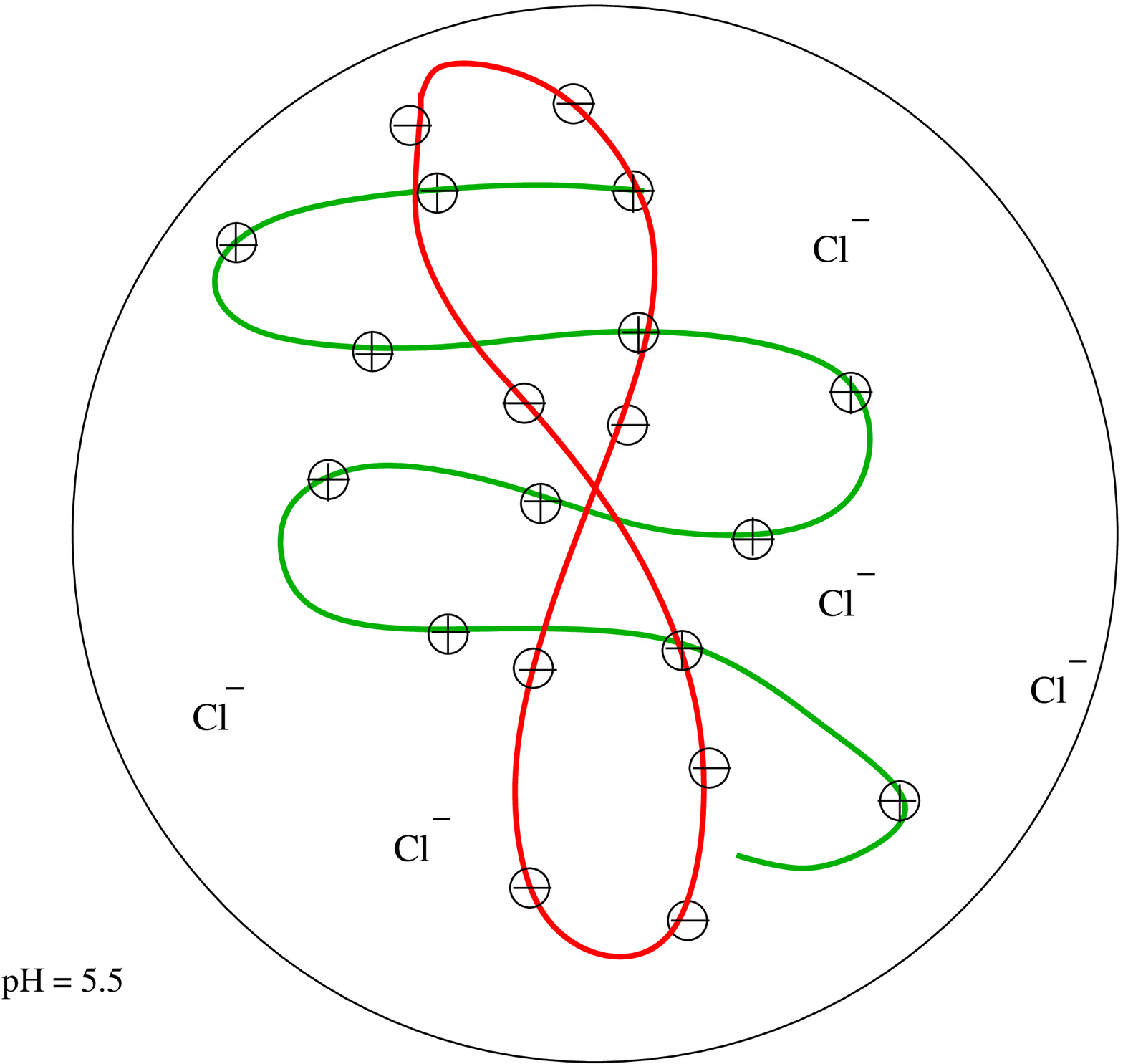}
\caption{The proton sponge hypothesis: protons $H^+$ and chloride ions $Cl^-$ enter into the endosome.  The increased $Cl^-$ concentration in the endosome with respect to the cytosol causes  an osmotic pressure on the endosomal membrane which can induce its disruption. Swelling mechanism: Protons bind to the polycation thus increasing its charge density and stiffness due to higher Coulomb repulsion between monomers. In the figure, the darker chain represents the DNA while the lighter one represents the polycation molecule. }
\label{mechanism:fig}
\end{figure}
In order to transfer DNA cargo to the nucleus, polyplexes must
escape from endosomes. This transmembrane mechanism or endosomolytic
process remains unclear and it is believed that high gene transfection
efficiency observed with PEI is due to its ability to avoid endosome
acidification and to promote complexe's release through a mechanism
called "proton sponge". This mechanism can be summarized as follow
(\ref{mechanism:fig}): the PEI contains amine groups, which can be
protonated at low pH. Thus due to an influx of chloride ions $Cl^{-}$
to maintain the overall electroneutrality, and water molecules, the endosome starts
swelling leading to an increased osmotic pressure, a destabilization
of the endosome and ultimately disruption. It has not been
proved yet whether {the} stress produced by the proton sponge effect and
by the swelling are the only mechanisms responsible for the
endosomal membrane disruption. However, recent experimental evidences argue
in favor of this escape mechanism \cite{Sonawane}. There are other possible mechanisms responsible for endosomal disruption such as swelling of the endosome due to internal charge repulsion triggered by an increased of ion
concentration in the endosome. This effect implies a direct physical interaction between the DNA-PEI
complex and the endosomal membrane.  The PEI-DNA complex can escape an endosome through a
membrane hole, probably due to direct interaction of polymers with
membrane  \cite{Bieber}. For example, the interaction between the
cationic polymer PEI and cellular components \cite{Helander1,Helander2} lead to Gram-negative bacteria
membrane disruption by free PEI, \cite{Oku}. Furthermore, PEI also
causes a liposomal membrane permeability to increase \cite{Klemm}.
Thus, PEI is capable of destabilizing lysosomes and this suggests
that it enables the DNA conjugated to the PEI to escape into the
cytoplasm. Finally, a wide variety of nanoparticles, including
different cationic polymers can physically disrupt lipid membranes
by forming nanoscale hole and membrane thinning \cite{Leroueil}.
Cationic polymers contribute to expand existing defects or can
directly induce small holes. Today, most common theories of pore
formation are derived from the classical nucleation theory
\cite{nucleation}, which unfortunately does not predict stable pores
of finite size. However long-lived pores that remained open for
several seconds have been observed \cite{pore1,pore2,pore3,pore4,pore5}.
Several theoretical models have succeeded to show that after pores nucleation, their enlargement
is expected to relax the surface tension, leading to the formation of a stable or
long-lived metastable pore \cite{stable-pore1,stable-pore2,stable-pore3,stable-pore4,stable-pore5,stable-pore6}.

For a planar membrane, increasing the lipid density occurs
concurrently to pore dilation, that reduces mechanical tensions.
After a vesicle opens, its internal content escapes, reducing the
osmotic pressure and the associated surface (Laplace) tension.
Although membrane fluctuation was recently modeled
\cite{FARAGO,membra-fluct,membra-fluct1}, there is still no consistent theory of pore formation
and in particular, it has yet to be elucidated whether polymer-membrane interactions and hole
formation are responsible for endosomal escape and how such process can be controlled to optimize gene cytoplasmic delivery.

\subsubsection*{Molecular simulation of DNA complexation in an endosome.}
To check whether protonation increased of a PEI chain is associated
with endosomal swelling disruption, we summarize here some recent findings
\cite{amoruso-holcman} about complexation between two oppositely charged polymers.
Other theoretical studies \cite{Winkler,Muthukumar} have shown that after complexation,
extended oppositely charged polymer chains collapse into a compact globule, even though the initial chain
configuration contained  bound ions.  Ions are progressively released by polycation-polyanion pairs
attraction and the final polymer complex, containing two chains separated from the rest of the ions.

In the presence of many interacting ions, the dynamics of a DNA molecule mixed with polycations is quite complex. But motivated by experimental data \cite{Choosako} suggesting that the number of polycations can affect the PEI-DNA interaction, we recently used Brownian simulations \cite{amoruso-holcman} to study the effect of various proton
concentrations on this interaction. We modeled polymers as flexible bead-spring chains, where each bead
represents a charged monomer, while ions are modeled as charged particles. The solvent is treated as a dielectric continuum with dielectric a constant $\epsilon=80$. We started with two oppositely charged chains whose monomer positions are  respectively
\beq
\x = \left(\x_1,\cdots,\x_N \right),\\
\y = \left(\y_1,\cdots,\y_M \right),
\eeq
mixed with ions at position
\beq
\z = \left(\z_1,\cdots,\z_P
\right),
\eeq
where the charge electro neutrality is preserved $P = M+N$. The
motion of a polymer chain in an overdamped medium is described by
the Smoluchowski's limit of the Langevin equation, which reduces to
a system of first order stochastic differential equations. The
dynamics of bead $i$ at position $\x_i$ in a potential $U(\x,\y,\z)$
is
\beq \label{eqfdt}
\dot{\x}_i + \frac{1}{\gamma}\frac{\p U}{\p \x_i } =  \sqrt{2D}\dot{\w}_i \hbox{ for }
i=1..N,
\eeq
where $D$ is the diffusion coefficient and $\dot{\w}_1,..\dot{\w}_N$
are N-independent $3-$dimensional Brownian motions and $\gamma$ is
the viscosity coefficient. The total potential $U = U(\x,\y,\z)$ is
the sum of various potential terms
\beq
U(\x,\y,\z) = U_{EL}(\x) +   U_{EL}(\y) + U_{bend}(\x) +  U_{bend}(\y)  + U_{LJ}(\x,\y,\z) + U_C(\x,\y,\z)
\label{pot}
\eeq
where
\begin{itemize}
\item   $U_{EL}(\x)$  is the polymer elastic energy: the potential well $U_k$
generated by two adjacent springs on the $k$-bead is the sum of the
two neighboring potentials:
\beqq
U_{k k+1}(\x_k,\x_{k+1})=k\left(\frac{1}{2} |\x_k
-\x_{k+1}|^2-l_0 |\x_k -\x_{k+1}|\right
) \\
U_{k k-1}(\x_k,\x_{k-1})=k\left(\frac{1}{2} |\x_k
-\x_{k-1}|^2-l_0 |\x_k -\x_{k-1}| \right),
\eeqq
where $k$ is the linear elasticity constant and $l_0$ the
equilibrium length. For $0<k<N$, the potential $U_k$ is
\beqq
U_k(\x_{k-1},\x_k,\x_{k+1})&=&U_{k k+1}(\x_k,\x_{k+1})+U_{k
k-1}(\x_k,\x_{k-1}) \\
U_0(\x_1,\x_{2})&=&U_{1 2}(\x_1,\x_{2})\\
U_N(\x_{N-1},\x_N)&=&U_{N N-1}(\x_N,\x_{N-1})
\eeqq
and the total potential is given by
\beq
U_{EL}(\x)= \sum_{k=1}^N
U_k(\x_{k-1},\x_k,\x_{k+1})
\eeq
The elastic energy $U_{EL}(\y)$ of the second chain is computed
similarly.
\item $U_{LJ}(\x)$ is the repulsive Lennard-Jones (LJ) potential, which models the excluded volume between all
the particles in the system (monomers and ions) and it is computed
as follows: the LJ-potential between two monomers of the same chain
is given by
\beq
  U_{LJ}(\x_i,\x_j)  = \left\{ \begin{array}{ll}
   4\epsilon_{LJ} \biggl[(\frac{\sigma}{\x_{ij}})^{12} - 2(\frac{\sigma}{\x_{ij}})^{6}\biggr]+ \epsilon_{L
J}   & \x_{ij} \leq \x_c \\
    0   &  \x_{ij}  > \x_c \\
\label{LJ:eq}
 \end{array} \right.
\eeq
where $\x_{ij} = |\x_i-\x_j|$ is the distance between the monomers,
$\epsilon$ is the depth of the potential well, $\sigma$ is the (finite)
 distance at which the inter-particle potential is zero. The LJ-potential is cut off at a
distance $\x_c = 2^{1/6}\sigma$ and the constant
$\epsilon_{LJ}$ is added to avoid discontinuity at $\x_c$.

A similar potential is used to compute the LJ-interaction between
monomers of the second chain  $U_{LJ}(\y_i,\y_j)$, monomers
belonging to different chains $U_{LJ}(\x_i,\y_j)$, monomers and
ions, $U_{LJ}(\x_i,\z_j)$ and  $U_{LJ}(\y_i,\z_j)$, ion-ion
$U_{LJ}(\z_i,\z_j)$, so that the total LJ potential is given by:
\beq
U_{LJ}(\x,\y,\z) = \sum_{i \neq j}U_{LJ}(\x_i,\x_j) +  \sum_{i,j}U_{LJ}(\x_i,\y_j) + \sum_{i,j}U_{LJ}(\x_i,\z_j) + \\
+ \sum_{i \neq j}U_{LJ}(\y_i,\y_j) +  \sum_{i,j}U_{LJ}(\y_i,\z_j) +
\sum_{i \neq j}U_{LJ}(\z_i,\z_j)
\eeq
\item The stiffness of a polymer is described by a bending energy of the form
\beq
U_{bend}(\x) = \frac{\kappa_{ang}}{2}\sum_{i=2}^{N}(\x_{i-1} - 2\x_i + \x_{i+i})^2 = \kappa_{ang}\sum_{i=1}^{N-1}(1-\mb{u}_i \cdot \mb{u}_{i+1}),
\eeq
where $\mb{u_i}= ( \x_{i+1} - \x_i ) / |\x_{i+1} - \x_i| $ is the
unit bond vector connecting two consecutive monomers and $\x_i$ is
the position of the $i-th$ monomer. This potential depends on the
angle $ \theta_i $ between two successive monomers since $\mb{u}_i
\cdot \mb{u}_{i+1} = cos\theta_i$ . The constant $\kappa_{ang}$ measures the bending
rigidity which is related to the persistence length of the polymer
by $l_p = \kappa_{ang}/(k_BT)$, where $k_B$ is the Boltzmann
constant and $T$ is the temperature \cite{Landau}. The persistence length characterize the stiffness of a
polymer and it is defined as follows:
\beq
\langle \cos \theta \rangle \sim \ds{\exp{\frac{l}{l_P}}}
\eeq
where $\theta$ is the angle between two vectors tangent to the
polymer separated by a distance $l$, and the cosines of the angles is averaged (angled brackets) over
all configurations. The stiffness energy $U_{bend}(\y)$ of the second chain is similarly computed.
\item The electrostatic potential between two charged monomers (charges are relocalized at the monomer) at
position $\x_i$ and $\x_j$ is given by the Coulomb potential
\beq
U_{C}(\x_i,\x_j) = \frac{e^2Z_iZ_j}{4\pi \epsilon} \sum_{i\neq j}\frac{1}{  |\x_i-\x_j|}
\label{Eq:COULOMB}
\eeq
where  $\epsilon$ is the dielectric constant of the medium and $Z_i$ the valence of the ion
$i$. The total electrostatic potential is
\beq
U_C(\x,\y,\z) = \sum_{i \neq j}U_C(\x_i,\x_j) +  \sum_{i,j}U_C(\x_i,\y_j) +  \sum_{i,j}U_C(\x_i,\z_j)+\\
+  \sum_{i \neq j}U_C(\y_i,\y_j) +  \sum_{i,j}U_C(\y_i,\z_j) +
\sum_{i\neq j}U_C(\z_i,\z_j),
\eeq
which represents the total Coulomb interactions between monomers on
a single chain, $U_C(\x_i,\x_j)$ and $U_C(\y_i,\y_j)$. For monomers
located on different chains  the potential is $U_C(\x_i,\y_j)$, for
monomer-ion interaction, it is  $U_C(\y_i,\z_j)$ and
$U_C(\y_i,\z_j)$, and finally the ion-ion interaction is given by
$U_C(\z_i,\z_j)$.
\end{itemize}
The dynamics of the second polymer with $M$ beads and positions $\y
= (\y_1,\cdots,\y_M)$ is governed by
\beq
\dot{\y}_i + \frac{1}{\gamma}\frac{\p U}{\p \y_i } =  \sqrt{2D}\dot{\w}_i \hbox{ for }
i=1..M.
\eeq
Finally the ions with positions   $\z = (\z_1,\cdots,\z_M)$ satisfy
\beq
\dot{\z}_i + \frac{1}{\gamma}\frac{\p U}{\p \z_i } =  \sqrt{2D}\dot{\w}_i \hbox{ for } i=1..P.
\eeq
The previous stochastic equations were solved numerically \cite{amoruso-holcman}, by standard
Euler's scheme. The results of the simulations are presented in the next section.

\subsubsection*{Numerical simulations of DNA-polymer interactions}
To check whether increasing the proton concentration modulates the DNA-PEI interaction
and leads to the deformation of the DNA-polycation complex, we ran various Brownian simulations
\cite{amoruso-holcman}, that we summarize now. We initially placed two oppositely
charged chains of total length $L_p$ in a 3D-cube of length $L$ and each monomer
position is randomly chosen at equilibrium (the two spherical angles
between monomers are uniformly distributed in the 2-sphere $S(2)$ of radius $l_0$). To
ensure electroneutrality, we added a number of ions, equal to the
difference between the DNA and polycation charges. Periodic boundary
conditions are imposed on the cube. $L$ was chosen such
that the density $\rho =(L_p/L)^3\approx 10^{-4}$ is small. In that
regime, there are no self-interactions between polymers. In a first
transient regime, we waited for each chain to relax to
equilibrium. Then, we move the chains close enough and
restart the simulations to evaluate their interaction. We found that the structure of the final
complex made of the two chains, after equilibrium is that of a flexible polycation winding around the DNA chain (Fig.\ref{twochains:fig}) and ions are uniformly distributed over the simulation box.
To further study the consequences of decreasing the pH and changing the PEI ionization
(accounting for the experimental procedure \cite{BEHR}), we repeated the previous simulations
in the presence of different ionic concentrations (Fig. \ref{twochains:fig}).

We further run additional simulations in a reflecting sphere, modeling an endosomal compartment to estimate the distribution of charges along the PEI monomers. We used \cite{amoruso-holcman} a Metropolis Monte-Carlo simulation \cite{BEROZA-PNAS} to evaluate the protonation configuration of a polymer having $N$ monomers. Each monomer is capable  of binding a proton. To estimate this distribution, we compute the free energy at equilibrium by finding
the minimum of
 \beq
G({\bf x}) = \sum_i^N \left(x_i k_BT \log 10 (pH-pK_i^0)\right) + \frac{1}{2}\sum_{i \neq j} W_{ij}q_iq_j ,
\label{gibbs}
\eeq
where ${\bf x}  = \{x_1,x_2,\cdots,x_N \}$ and $x_i = 1$ when site $i$ is protonated, $0$ otherwise.
The first term in \ref{gibbs} is the  energy required to protonate a site,
depending on $\epsilon$, the pH and the $pK$ of the PEI \cite{BEROZA-PNAS} and $W_{ij}$ is the electrostatic Coulomb interaction energy  between sites $i$ and $j$ when they are both charged.
Thus, the average protonation of a given site $x_i$ is obtained by averaging (Boltzmann
weighted sum) over all configurations:
\beq
\langle x_i \rangle = \frac{\sum_{x_i}x_i\exp^{-G(x)/k_BT}}{\sum_{x_i}
\exp^{-G(x)/k_BT}}
\label{average}
\eeq
We used \cite{amoruso-holcman} the Metropolis algorithm
\cite{Metropolis} to sample the most probable states of $x_i$ and
Fig.\ref{titration_DD} summarizes the configuration distributions
for various pH.
\begin{figure} [htb]
\includegraphics[angle=0,width=0.55\textwidth]{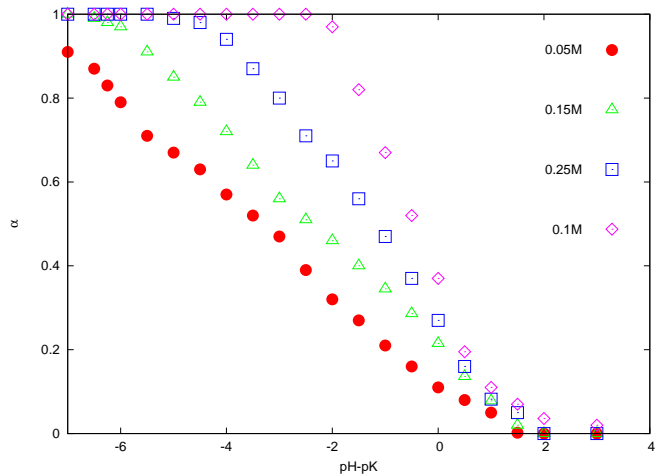}
\caption{{\bf Simulated protonation curves of PEI  for different salt concentrations},
0.01M,0.05M, 1.0M.  At high salt concentration, there is a steep
rise in the protonation as the pH approaches $pK_0$ and the chain is
fully protonated for $pH-pK_0 \sim 2$. As the salt concentration
decreases, protonation of the chain becomes more difficult, because
there are fewer ions to screen the electrostatic repulsion of
charges on the chain.}
\label{titration_DD}
\end{figure}
At physiological concentration $C_s = 0.15M$ and  $pH = 7.4$, then
$pH-pK \simeq 2.7$, that is $50\%$ of monomers are protonated,
higher than experimental measurements ($20\%$ of protonated
sites \cite{Suh}). DNA-PEI complex seems to evolve to different
configurations depending on the pH. In Fig. \ref{twochains:fig}a,
the DNA-polycation forms a complex. The polycation is only partially
charged (indeed $20\%$ of the amine groups are protonated
\cite{Suh}). The polycation behaves almost like a flexible chain, while DNA monomers are subject
to strongest Coulomb repulsion making it stiffer and more extended,
and almost immobile during complexation.
\begin{figure} [htb]
\includegraphics[angle=0,width=0.28\textwidth]{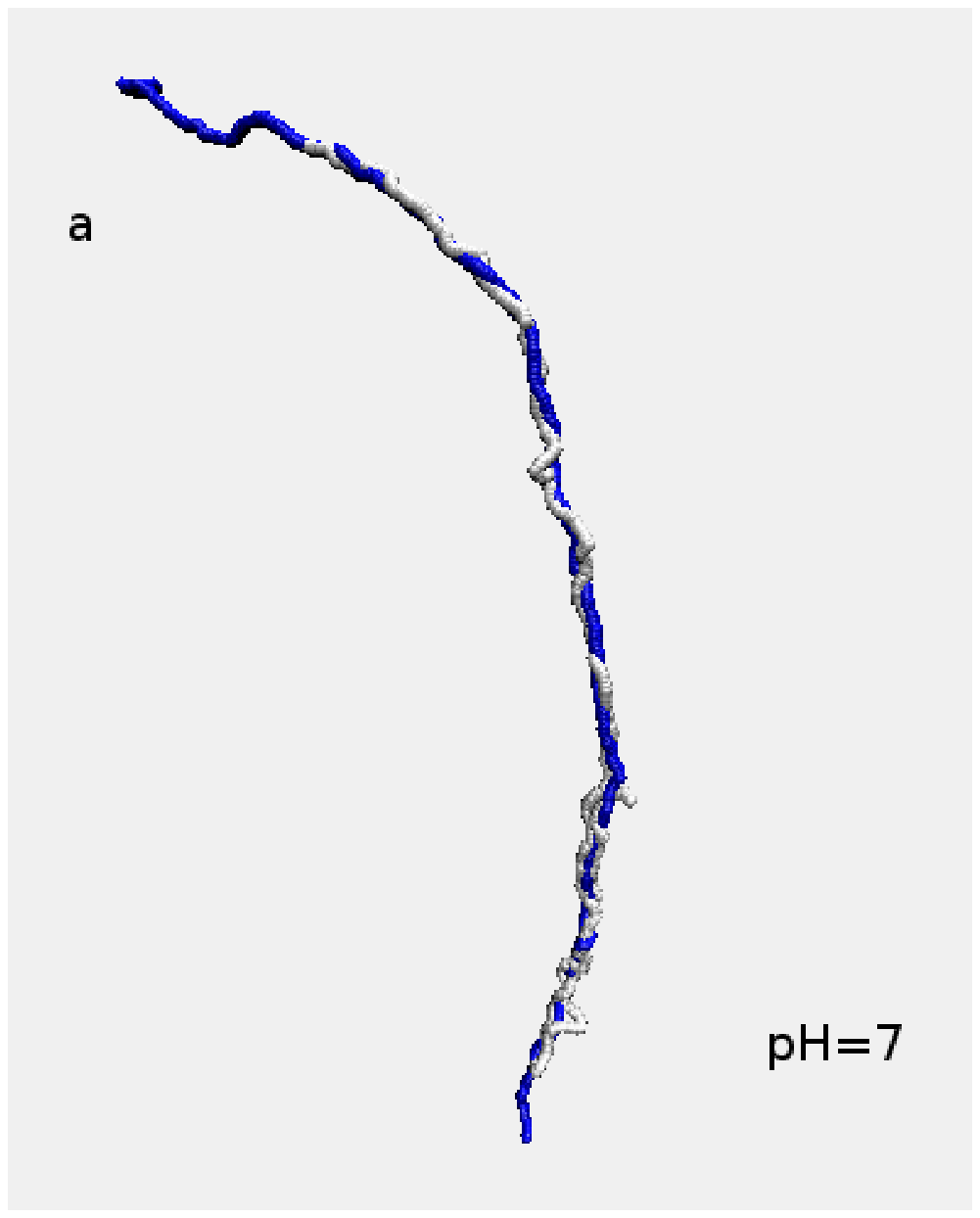}
\includegraphics[angle=0,width=0.28\textwidth]{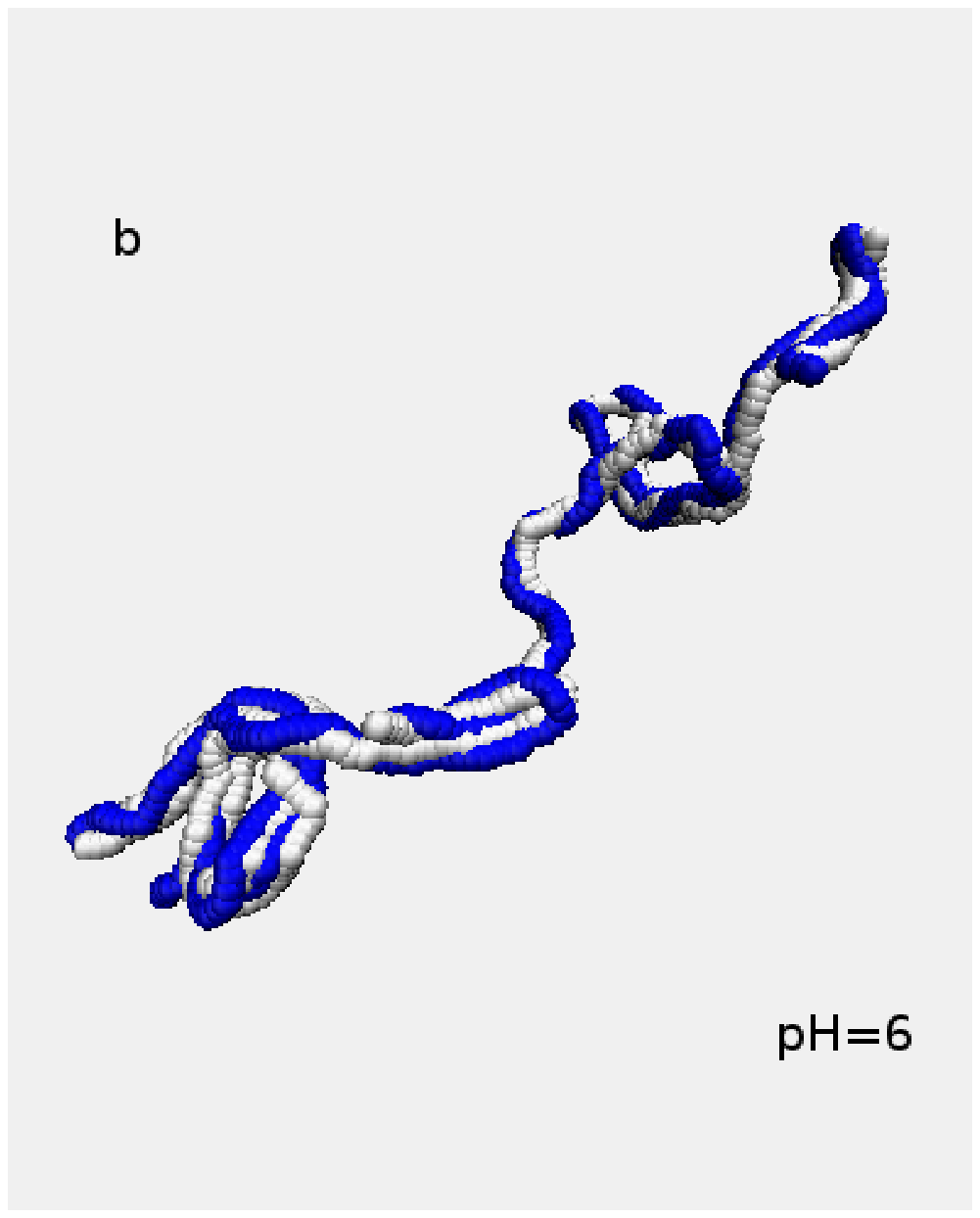}
\includegraphics[angle=0,width=0.28\textwidth]{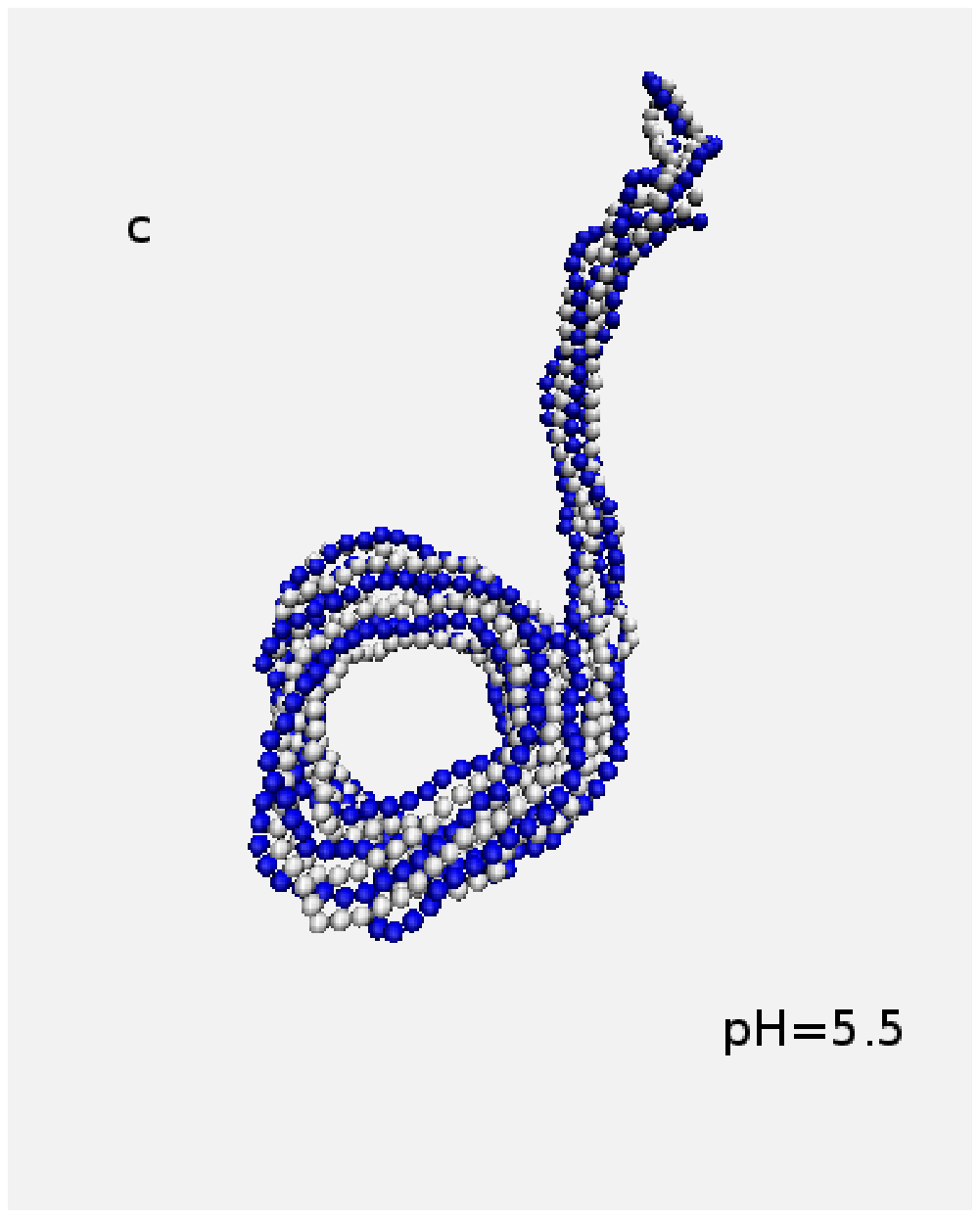}
\caption{{\bf Snapshots of the equilibrium configurations of the DNA/PEI
complex} at pH = 7.0 (left), pH = 6.0 (middle), ph = 5.5 (right).
During the acidification process the size of the complex decreases.
Each image represents the configuration at fixed pH after $10^6 -
10^7$ steps of the Brownian dynamics, which corresponds roughly to
$1-10s$.}
\label{twochains:fig}
\end{figure}
Fig. \ref{twochains:fig}b and \ref{twochains:fig}c show the DNA/PEI complex
configurations respectively at pH = 6 and 5, during the
endosomal acidification. Increased protonation could induce swelling
\cite{BEHR} of the complex and consequently promoting endosome
rupture. The result in \cite{amoruso-holcman} predicts that
increasing attraction between oppositely charged polyelectrolytes
lead to a more compact complex structure. Upon increasing the protonation
state of the PEI, the positively charged chain condenses more
tightly on the DNA, which become more flexible because
its charged are now screened. The resulting complex resembles a typical
toroid configuration found for DNA interacting with multivalent ions
\cite{Stevens}. The same polyplex structure was previously obtained
\cite{YANG-MAY} by solving numerically modified Poisson-Boltzmann
equations for the electrostatic potential and the polymer
concentration. These results \cite{amoruso-holcman} suggest  that two oppositely
charged polymers condense upon lowering pH and thus this mechanism cannot be responsible for the swelling of the
endosome and endosomal membrane disruption.
\begin{figure} [htb]
\includegraphics[angle=0,width=0.28\textwidth]{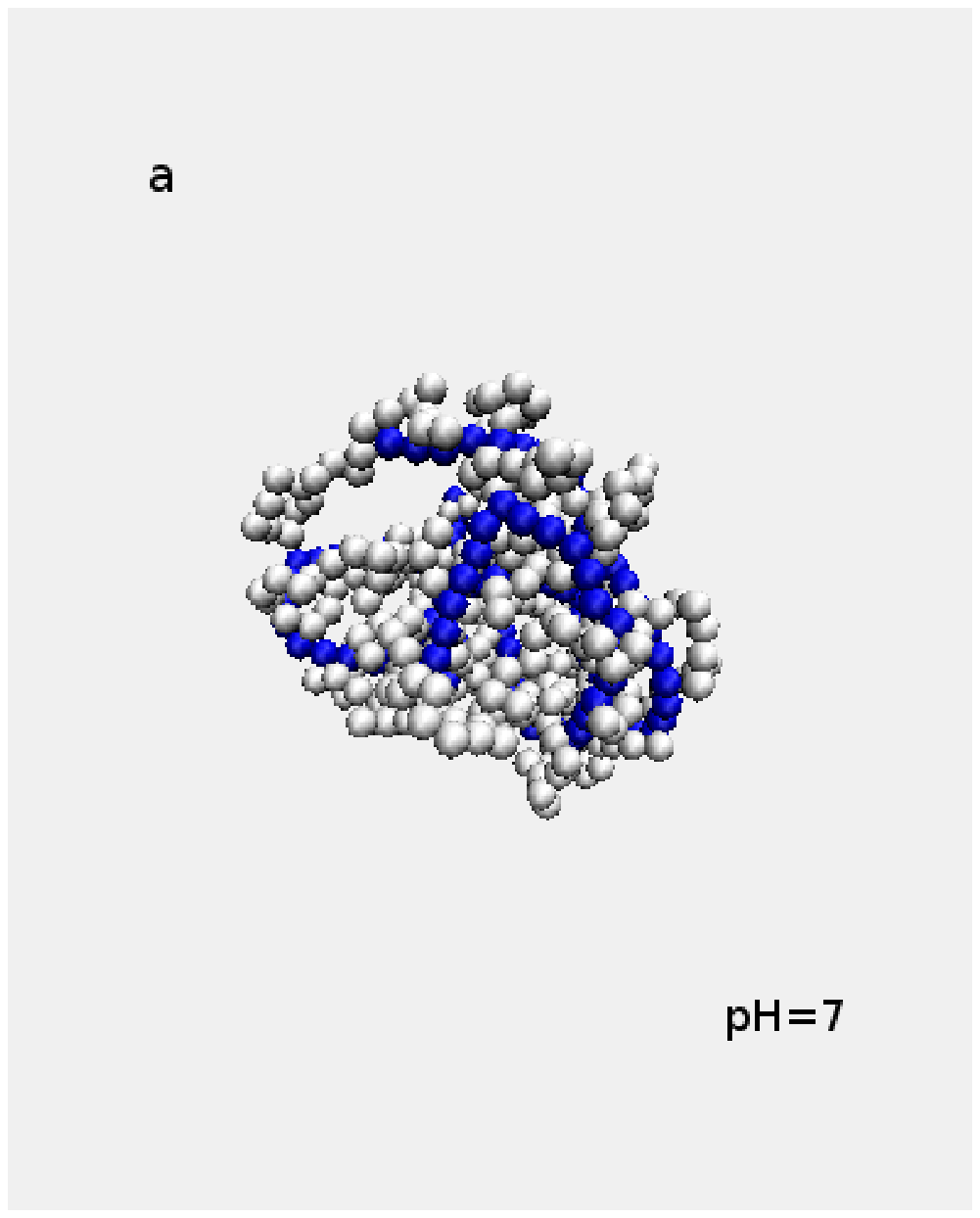}
\includegraphics[angle=0,width=0.28\textwidth]{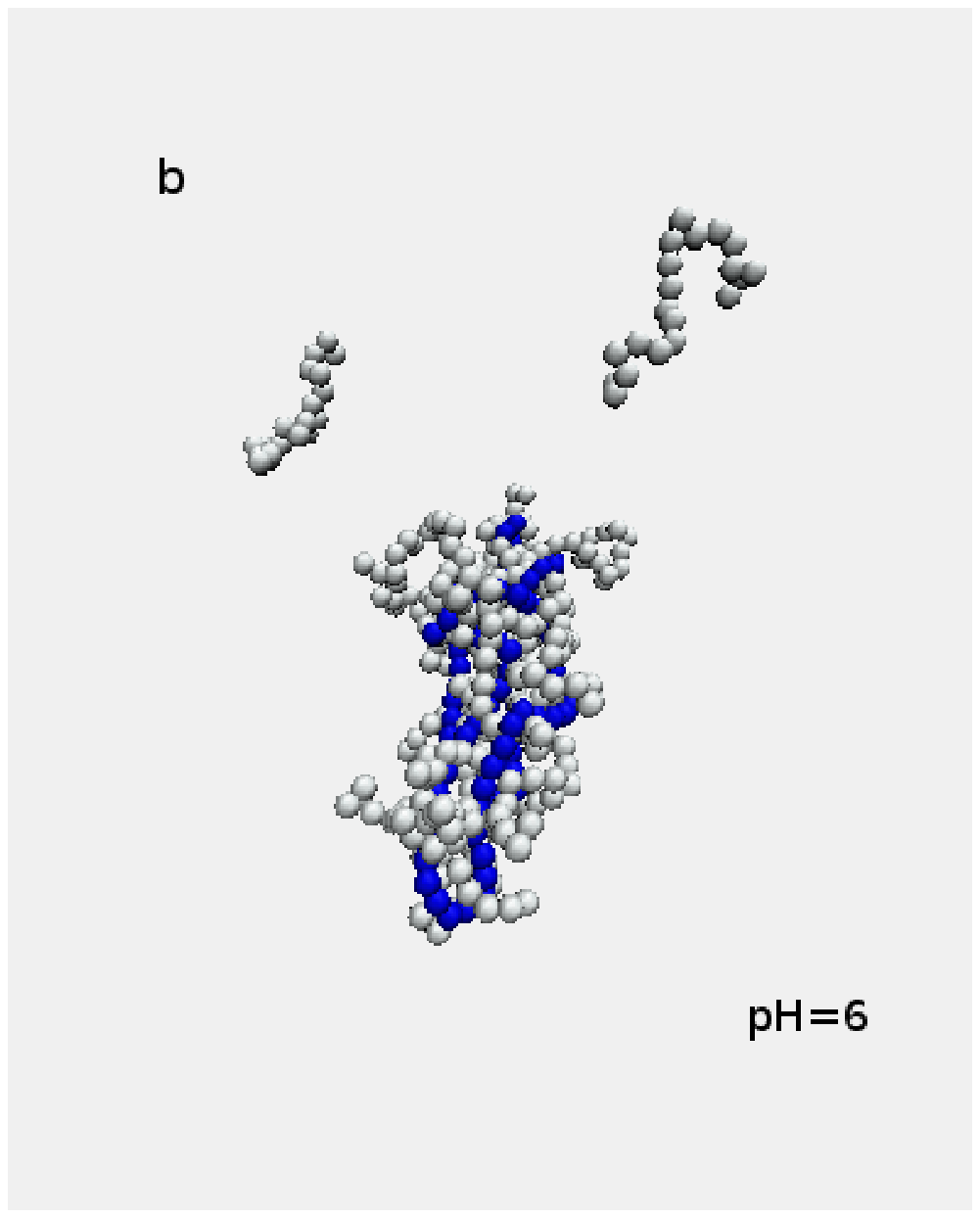}
\includegraphics[angle=0,width=0.28\textwidth]{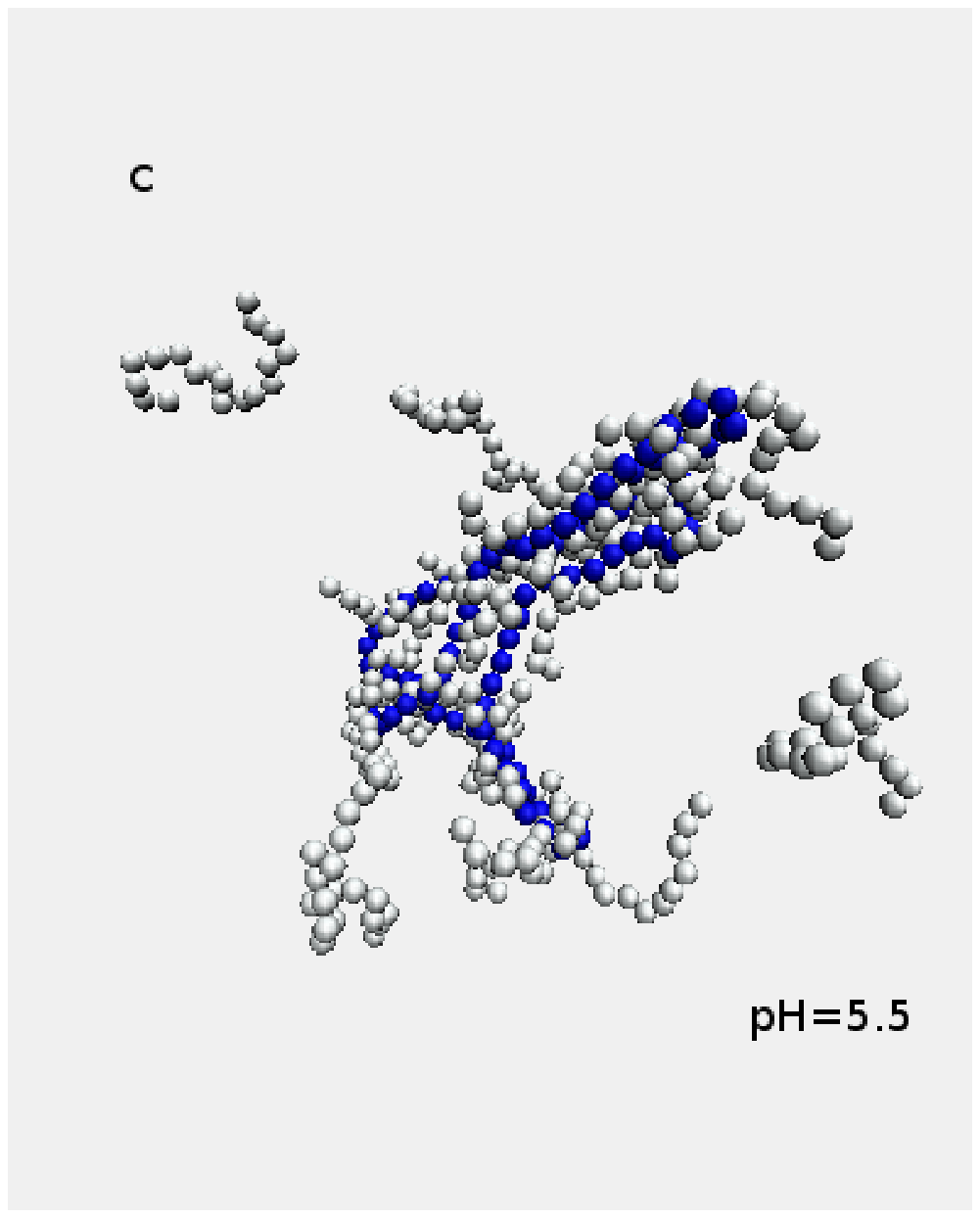}
\caption{ Snapshots of the equilibrium configurations for a DNA molecule complex with many small polycations,
pH = 7.0 (left), pH = 6.0 (middle), ph =5.5 (right). During the
acidification process, the complex-DNA can swell due to an increased
Coulomb repulsion and consequently some polycations can be released
from the complex. Each frame represents the configuration at fixed
pH after $10^6 - 10^7$ steps of the Brownian dynamics.}
\label{many:fig}
\end{figure}
We shall now review another simulation sets, aimed to better characterize
an experimental procedure where the DNA molecule is mixed in solution
with a fixed concentration of PEI,  where the ratio $r=N/P$ of the
N amine to P phosphate groups is fixed. A value of $N/P = 4 - 10$ has been reported from experiments \cite{Choosako} and we simulate a DNA molecule with $N_1=100$ monomers when it is complexed with
$20$ polyactions, each of length $N_2 = 20$, making  the ratio $r=N/P
= 4$. We use the same numerical method as described above with many polycations.
Fig. \ref{many:fig}a shows the equilibrium configuration of the DNA
complexed with the small polycations at physiological pH with
periodic boundary conditions, while Fig.\ref{many:fig}b and
\ref{many:fig}c represent the influence of acidification in a
confined spherical microdomain. At physiological pH, the DNA complex is less
compacted and more flexible due to an excess of positive charges resulting in additional screening, compared with the situation of two chains only (Fig \ref{twochains:fig}a). During
acidification, at higher charge fraction of the PEI, some
polycations are released from the complex and are free to diffuse
in the endosome. The presence of free polycations has been also
observed experimentally \cite{CLAMME}.
Recent experiments \cite{Boeckle} have shown that purified
polyplexes without free PEIs were less efficient in transfection
compared to non-purified polyplexes.  Hence interaction of free PEIs with the endosomal membrane,
could play a role in endosomal escape, however a detailed model
of membrane destabilizaton due to membrane-polymer interaction is currently missing.

In summary, Brownian simulations can be used to study processes
occurring at intermediate scale between molecular and cellular. We
have reviewed here some approaches  method to simulate the interaction of
oppositely charged complex polymers in endosomal following
protonation. We reported that surprisingly, the protons and chloride
endosomal influx which lead to an increase in the ionic
concentrations and the enlargement of the polycation-DNA complex
(due to internal charge repulsion), in the range of parameters such
as $pH \in [5.5-7]$, the number of poycations from 1 to 20
\cite{amoruso-holcman}, seems not to have any direct role in the
endosome disruption. This result suggests that more refined scenario should be considered to study
the direct PEI-DNA interaction during endosomal membrane disruption.
\section{Modeling the endosomal step of viral infection}
To better understand how viruses can be used as gene vectors, we
now focus on one of the fundamental step of early infection, where the viral particles travel inside an endosome.
Indeed, most viruses enter cells in an endosomal compartment, after binding to
specific membrane receptors. To undergo cytoplasmic or nuclear
replication \cite{Arhel,Zhuangreview}, and to avoid degradation in
acidic lysosomes, viruses must then successfully escape the
endosome. Enveloped viruses, such as Influenza, contain membrane-associated glycoproteins mediate the fusion between the viral and endosomal membranes. In particular, acidification of the
endosome triggers the conformational change of the influenza
hemagglutinins (HA) into a fusogenic state, leading to endosome-virus
membranes fusion and genes release inside the cytoplasm.

Developing a biophysical model of the influenza endosomal step offers
a general framework to study the influence of several parameters such as the endosomal size, the
number of viral particles or the molecular structure of
glycoproteins on the efficiency of viral escape. In particular, fusogenic peptides derived from viral
glycoproteins are increasingly used in cationic synthetic vectors
\cite{vector1,vector2}, and the pH-sensitivity of fusogenic
glycoproteins can now be tuned by modifying the electrostatic
stability of the fusogenic complex \cite{herrmann2}. These
engineered glycoproteins shall serve to design efficient gene
vectors and quantitative models will help optimizing glycoprotein
molecular properties with respect to the endosomal escape efficacy of
the vector.

We shall review previous models \cite{lagache-env} aimed to estimate the residence time of a viral particle inside an endosomal compartment.
These models are based on considering the accumulation of discrete proton binding events, leading to the
conformational change of HAs. This is the limiting step of genes
release in the cytoplasm and the model was built into two steps:
first, by fixing the concentration of protons, we used a Markov
jump analysis to estimate the mean time for protons to bind the HA binding sites
until a threshold is reached, triggering its conformational change into
a fusogenic state. Interestingly, our analysis allow to extract from the HA
conformational change kinetics measured experimentally at different
pH \cite{kinetics} the binding rates and the number of bound sites that are needed for the protein to change conformation. Second, coupling the pH-dependent
conformational change of HA glycoproteins with a linear proton influx rate, we analyzed the endosomal escape dynamics of influenza
viruses. We predicted \cite{lagache-env} that the size of the
endosome drastically impacts both the escape kinetics and pH, which
reconciles different experimental observations:  while a virus can
escape from small endosomes (radius of $80nm$) in the cell periphery
at a pH $\sim 6$ in about $10$ minutes \cite{sakai}, it can also
be routed towards the nuclear periphery, where escape from larger endosomes (radius of $400 nm$) is rapid
(less than one minute)  at pH $5$ \cite{zhuang}.  We shall
now review in details the modeling of these two steps.

\subsubsection*{Modeling the conformational change of glycoproteins}
To estimate the conformational change rate of a single HA
glycoprotein at a given proton concentration $c$, we considered
\cite{lagache-env} that the protein changes conformation
instantaneously when the number of bound sites reaches a critical
threshold $n_{crit} \leq n_s$, where $n_s$ is the total number of HA binding
sites.

To follow the time dependent number of protonated sites, we use a stochastic analysis based on
Markov jump processes \cite{KM1,KM2,KM3,schuss}, where during time $t$ and $t+\Delta t$, the amount of protonated sites
$X(t,c)$ can either increase with a probability $r(X,c)\Delta t$
when a proton binds to a free site, decreases with probability
$l(X,c)\Delta t$ when a proton unbinds or remains unchanged with
probability $1-l(X,c)\Delta t-r(X,c)\Delta t$. Using a scaled
variable $x(t,c)=\epsilon X(t,c)$ where $\epsilon=1/n_s$ and
$\Delta x=x(t+\Delta t,c)-x(t,c)$, the transition probabilities
satisfy
\begin{eqnarray*}
\Pr\{\Delta x=\epsilon|x(t,c)=x\}=r(x,c)\Delta t,\\
\Pr\{\Delta x=-\epsilon|x(t,c)=x\}=l(x,c)\Delta t,\\
\Pr\{\Delta x=0|x(t,c)=x\}=\left(1-r(x,c)-l(x,c)\right)\Delta t.
\end{eqnarray*}
For a fixed proton concentration $c$, the transition probability function $p(y,t|x,c)$ that the proportion $x(t,c)$ of protonated sites is equal to $y$ at time $t$, $x(t,c)=y$, given that $x(0,c)=x$ is solution of the backward master equation \cite{KM1,KM2,KM3,schuss}:
\begin{eqnarray}
p(y,t|x,c)&=&p(y,t-\Delta t|x+\epsilon,c)r(x,c)\Delta t
+p(y,t-\Delta t|x-\epsilon,c)l(x,c)\Delta t \nonumber \\&+&
p(y,t-\Delta t|x,c)(1-r(x,c)\Delta t-l(x,c)\Delta t),
\end{eqnarray}
which can be expanded as (Kramers-Moyal expansion)
\begin{eqnarray}
\frac{ \partial p}{\partial t} &=&L_x
p=r(x,c)\sum_{n=1}^{\infty}\frac{\epsilon^n}{n!}\left(\partial_x\right)^n
p(y,t|x,c) \nonumber \\ &+&
l(x,c)\sum_{n=1}^{\infty}\frac{\left(-\epsilon\right)^n}{n!}\left(\partial_x\right)^n
p(y,t|x,c).\label{BKME}
\end{eqnarray}
{ The first time $\tau$ a glycoprotein is filled up to a critical threshold $x_{crit}=n_{crit}/n_s$ is the first passage time for the bound protons $x(t,c)$ to reach the level $x_{crit}$. The mean first passage time $\tau(x,c)$ is defined as the conditional expectation $\tau(x,c)=E[\tau|x(t=0,c)=x]$, and satisfies \cite{schuss,VanKampen}:
\begin{eqnarray*}
&&L_x \tau(x,c)=-1 \hbox{ for }x \hbox{ in }[0,x_{crit}],\\
&&\tau(x,c)=0 \hbox{ for } x=x_{crit} \hbox{ and }\frac{\partial \tau(x,c)}{\partial x}= 0 \hbox{ for }x=0.
\end{eqnarray*}
We approximate the conformational change mean time with
$\tau_0(c)=\tau(x_0(c),c)$, where $0<x_0(c)<x_{crit}$ is the mean
number of bound protons (concentration $c$), the leading order in
$\epsilon\ll 1$ \cite{KM1,KM2,KM3,schuss} is
\beq\label{taucc}
\tau_0(c)\approx C(\epsilon,c)
\left(1-\left(\frac{l(x_{crit},c)}{r(x_{crit},c)}\right)^{\ds{-(x_{crit}-x_0(c))/\epsilon}}\right),
\eeq where \beqq C(\epsilon,c)\approx
\frac{1}{r\left(x_0(c),c\right)}\frac{\sqrt{\frac{2\pi}{\epsilon
\frac{d}{dx}\left(l/r\right)\left(x_0(c),c\right)}}}{\phi(x_{crit},c)}
\eeqq and \beqq
\phi(x,c)=\frac{l(x,c)/r(x,c)-1}{\sqrt{l(x,c)/r(x,c)}} \,e^{\ds{-\frac{1}{\epsilon}\int_{x_0(c)}^{x}log\left(l(s,c)/r(s,c)\right)
ds}} .
\eeqq
Formula (\ref{taucc}) relates the conformational change mean time of a single HA at a fixed protons concentration $c$  with the binding/unbinding rates $r(x,c)$ and $l(x,c)$ \cite{lagache-env}
\beq
r(x,c)=K c n_s(1-x), \hbox{ and } l(x,c)=l(x)=K n_s(1-x) 10^{-\left(3\left(1-x\right)+4\right)}, \label{rl}
\eeq
obtained from the mean number $x_0(c)$ of HA protonated sites at different pHs \cite{protonation}. In addition, $n_s=9$ is the number of binding sites \cite{protonation}, $c$ the concentration of free protons in the endosome and $K$ the proton binding rate to free binding sites. Using these transition rates in formula (\ref{taucc}), and comparing the
theoretical conformational change mean time $\tau_0(c)$ with
experimental data \cite{kinetics}, we found that $K\approx 7.5*10^3 L.mol^{-1}s^{-1}$ and $x_{crit}\approx 0.7$
\cite{lagache-env}. Interestingly, the theoretical curve $\tau_0(c)$ describes the entire range of experimental data set of the influenza hemagglutinin \cite{kinetics}, confirming the validity of our approach. Finally, comparing the mean proton binding time $\tau=1/(Kc)$ with the mean time
$\tau_d$ for a proton to find a binding site by diffusion \cite{holcmanschuss}, we found \cite{lagache-env} that:
\beq
\tau_d/\tau \approx 10^{-4},
\eeq
suggesting that the HA binding time is dominated by a high activation
barrier, which guarantees the stability of a conformational change, that cannot be easily
triggered randomly (table 2 \cite{kinetics}). The theory predicts that the  HA conformational changes occurs when roughly $x_{crit}n_s\approx 6$ HA binding sites are protonated.

\subsubsection*{Modeling the endosomal escape of the influenza virus.}
During endosomal maturation, protons enter actively through V-ATPase
pumps located in the endosomal membrane, leading to pH decrease and
the conformational change of HAs into a fusogenic state \cite{HA}.
After fusion of the viral and endosomal membrane, the
influenza genes can be released inside the cytoplasm.

For a linear time dependent proton influx $\lambda t$, the the rate $\lambda$ is proportional to the number of pumps and thus to the endosomal surface \cite{lagache-env}. Consequently, $\lambda$  may drastically increase as the endosome matures and increases its size by fusion of early endocytic vesicles into larger compartments \cite{zerial}. As protons accumulate into the endosome, they can bind to influenza HAs,
triggering their change of conformation when exactly $6$ sites are
protonated.

The entire viral membrane in contact with the endosome
seems to fuse before genes escape, as observed in electron
microscopy images (figure 5c \cite{sakai}). Furthermore, another step of endosomal escape consists in the enlargement of
the fusion pore, that should rely on the activation of additional HA
located nearby the contact zone, between the virus and the endosome
membranes \cite{leikina}. In previous models \cite{lagache-env},
we accounted for this complex cooperative mechanism between activated HAs nearby the contact zone, by considering
that genes are released in the cytoplasm when the total number of
activated HAs among the viral envelope reaches a threshold
$0\leq T \leq N_{HA}$, where $N_{HA}=400$ \cite{imai} is the total
number of HAs covering a single virus. Consequently, the escape
release time  $\tau_e$ is defined by
\beq
\tau_e=\inf\{t|HA_6(t)\geq T\}. \label{taue}
\eeq
where $HA_0(t), HA_1(t) \ldots HA_6(t)$ are the number of HAs that
have bound $0,1 \ldots 6$ protons at time $t$. The acidification time course of
an endosome containing an influenza virus is related to the number
of free protons located in the endosome by $P(t)=\mathcal{N} V_0 c(t)$ ,
where $c(t)$ is the associated endosomal concentration at time $t$.
Using the on-rate of a proton to a HA free binding site,
$\tilde{r}\left(x\right)=r\left(x,P(t)\right)/P(t)=K
n_s(1-x)/(\mathcal{N} V)$ and the off-rate $l(x)$ (\ref{rl}), the kinetics equations are
\beq
HA_0+P \overset{\tilde{r}(0/n_s)}{\underset {l(1/n_s)}{\rightleftharpoons}}  HA_1 \nonumber \\
HA_1+P \overset{\tilde{r}(1/n_s)}{\underset {l(2/n_s)}{\rightleftharpoons}}  HA_2\nonumber \\
\ldots\nonumber \\
HA_5+P \xrightarrow{\tilde{r}(5/n_s)} HA_6 \label{chemical}
\eeq
and the associated mass action law leads to the following differential
equation system
\beq
\frac{dP(t)}{dt}&=&\lambda-\sum_{i=0}^{5}\tilde{r}\left(\frac{i}{n_s}\right)P(t) HA_{i}(t)+\sum_{i=1}^{5}l\left(\frac{i}{n_s}\right)HA_{i}(t\nonumber)\\
\frac{dHA_0(t)}{dt}&=&-\tilde{r}\left(\frac{0}{n_s}\right)P(t)HA_{0}(t)+ l\left(\frac{1}{n_s}\right)HA_{1}(t)\nonumber\\
\frac{dHA_1(t)}{dt}&=&\left(\tilde{r}\left(\frac{0}{n_s}\right)HA_0(t)-\tilde{r}\left(\frac{1}{n_s}\right)HA_{1}(t)\right)P(t) + l\left(\frac{2}{n_s}\right)HA_{2}(t)-l\left(\frac{1}{n_s}\right)HA_{1}(t)\nonumber\\
\ldots\nonumber \\
\frac{dHA_6(t)}{dt}&=&\tilde{r}\left(\frac{5}{n_s}\right)HA_5(t) P(t), \label{differential}
\eeq
where the proton influx rate in the endosome is a linear function of
time $\lambda t$. The initial conditions ($t=0$) are a neutral
medium $pH=7$ and $ \ds{P_0= \mathcal{N} V_0 10^{-7}\approx 1}
\hbox{,  } HA_0(t=0)=N_{HA} \hbox{ and } HA_i(t=0)=0 \hbox{ for }
1\leq i \leq 6$.

The two critical unknown parameters of the model are the protons
influx rate $\lambda$ and the mean number $T$ of activated HAs,
needed for the large fusion pore formation allowing gene release inside
the cytoplasm. These parameters were obtained by solving numerically the system
of equations \ref{differential}, and by comparing the time $\tau_e$
(formula \ref{taue}) with the experimental mean escape time ($\approx 10$
minutes) obtained for the virus-endosome fusion in Hela cells
\cite{sakai}: we found \cite{lagache-env}
\beq
T \approx 50\% N_{HA}=200 \hbox{, and } \lambda \approx 3 s^{-1}.
\eeq
\subsubsection*{Using modeling approaches to study various endosomal pathways}
Viral fusion and genes release from small endosomes  (radius $\sim80$ nm) have been observed in cell periphery \cite{sakai}, starting only $10$ minutes after viral entry. Although the endosomal pH in these small endocytic vesicles is not precisely known, it is believed to be $\approx 6$. However, most viruses are routed towards the nuclear periphery into larger endosomes (radius $\sim 450nm$), where the escape time decreases to $1-2$ minutes and the associated escape pH to $\approx 5$ \cite{zhuang} (FIG. \ref{pathway}a).

To model the escape process in these two endosomal pathways, we developed a model \cite{lagache-env} that accounts for the endosomal size, where the proton influx rate is scaled to the endosomal surface $\lambda(r)= \left(r/r_0\right)^2 \lambda =\left(r/(80nm)\right)^2 3
s^{-1}$ \cite{lagache-env}. Taking this scaling into account, we found  by numerical resolution of equations (\ref{differential}) that the escape pH and mean escape time are drastically decreased as the endosome becomes larger. When the threshold to fusion is fixed to $T=200$ activated HAs  \cite{lagache-env}, the escape pH is $5.8$, and the escape time is $10$ minutes for an endosomal radius of $80nm$, while the escape time
and the pH drop to $40s$ and $5$ respectively for a radius of
$450nm$ (Fig. \ref{pathway}b-c).

In the general context of cytoplasmic trafficking, we proposed
\cite{lagache-env} that, when viruses use the periphery pathway where
they leave the endosome far away from the nucleus, at a pH $6$, they can further
escape the rapid inactivation process occurring at $pH<5.4$ \cite{inactivation}.
In that case, they still have to travel through the risky cytoplasm to a
small nuclear pore to deliver their genetic material. However,  for viruses
using the other pathway leading directly to the perinuclear area, the escape endosomal
$pH$ drops to $\leq 5$, thus this exposure  leads to possible HA degradation resulting in greatly diminishing the escape capacity.
Consequently, it is not clear which pathway is more efficient for viral gene delivery.
\begin{figure} [htb]
\includegraphics[angle=0,width=1\textwidth]{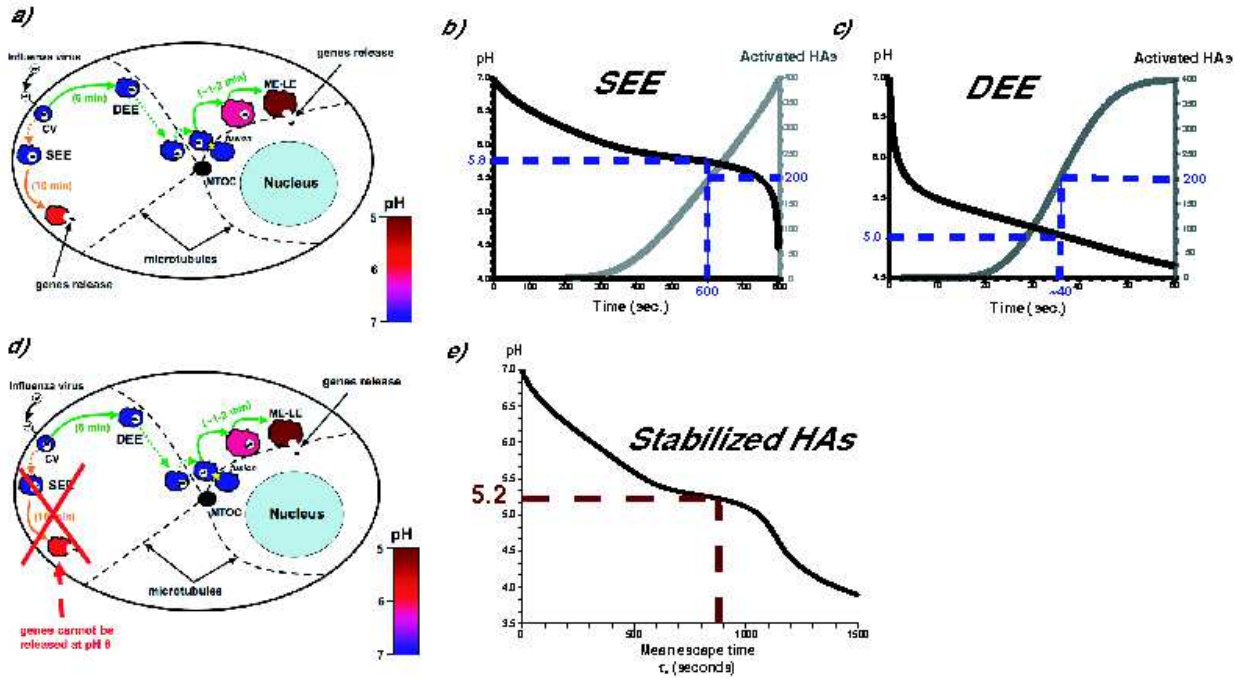}
\caption{ \textbf{Schematic representation of the influenza virus endosomal pathways. Genes are released in
the cytoplasm and the mean escape time and pH depends on the endosomal radius} \textbf{(a)}
After cell entry, the coated vesicle (CV) containing the virus can either be
routed towards small static early endosome (SEE) in cell periphery
\cite{zhuang-cell, sakai}, where fusion occurs after $10$ minutes at pH $~6$ \cite{sakai}(orange pathway), or it can
fuse with dynamic early endosomes (DEE) traveling along microtubules, to reach the perinuclear region and rapidly fuse from
larger endosomes  (green pathway) \cite{zhuang,zhuang-cell}. In the second case,
fusion time is decreased to $1-2$ minutes and the escape pH
drops to $~5$. \textbf{(b)} Time evolution of the number of
activated HAs $HA_6(t)$ (grey line) and the endosomal pH (solid
line) for a small SEE with $r=0.08\mu m$ \cite{sakai} (proton influx
rate $\lambda=3s^{-1}$). The pH is maintained above $5.8$ for
$HA_6(t) \leq 50\% n_{HA}=200$ (highlighted with blue dotted lines)
leading to genes release in the cytoplasm. \textbf{(c)}  For a
larger perinuclear DEE with $r=0.45 \mu m$ \cite{zerial}, the proton
influx rate has been scaled to the endosome surface
$\lambda=\left(45/8\right)^2 3s^{-1}=95s^{-1}$, which leads to a
rapid gene release in about $40$ seconds. d-e) {Engineered
Stabilized HAs with an increases low-pH stability shall prevent
viral vectors escape from peripherical endosomes at pH
$6$}.}\label{pathway}
\end{figure}
\subsubsection*{Perspectives in developing simulations to optimize synthetic virus-like vectors}
For many enveloped viruses, conformational change of glycoproteins
into a fusogenic conformation is triggered either by cumulative discrete
binding of protons (e.g. class I, II and III fusogenic glycoproteins
\cite{flavivirus,rhabdovirus}), or by low-pH activated proteases (e.g.
Ebola GP protein and the SARS coronavirus spike protein S
\cite{ebola,belouzard}). In addition, fusogenic glycoproteins are increasingly used in many
synthetic drug delivery systems such as siRNA delivery systems
\cite{sirna} or cationic synthetic gene vectors \cite{vector1,vector2}.
Thus, the generic model developed in  \cite{lagache-env} for the
influenza virus can be extended to this large class of viruses and synthetic vectors.
Furthermore, the pH stability of influenza glycoproteins has been recently tuned
by incorporating or removing electrostatic bonds into the fusogenic
complex \cite{herrmann2}. Consequently, developing biophysical models to quantify
the endosomal escape kinetics and pH of viruses with respect to
glycoproteins molecular properties is timely and could be further applied
 to optimize gene vector design. Indeed approaches developed in \cite{lagache-env}
 can predict the endosomal time course of vectors or mutant
viruses covered by these HA variants. In \cite{lagache-env}, we have
predicted that using stabilized forms of HA glycoproteins
\cite{herrmann2} shifts the escape pH in small endosomes from $6$ to $5$, which may prevent
vectors to use the peripherical endosomal pathway (figure \ref{pathway} d-e).
Consequently, if genes degradation inside the cytoplasm is a
limiting barrier to transfection efficiency, we predict that
stabilized mutant viruses may increase the ability of  vectors
to deliver genes to nuclear pores.

\section{Modeling gene carrier cytoplasmic dynamics}
We end this review by a section on quantifying the
success of cytoplasmic trafficking. Indeed, following the endosomal escape, viral and synthetic gene vectors have to travel through the
crowded and risky cytoplasm to reach the nucleus and deliver their genome through the nuclear pores. Recent imaging techniques now allow single particle tracking \cite{Seisenberger, Arhel,brandenburg-zhuang}, which can be used to model cytoplasmic motion of gene vectors. Because gene vectors do not possess means of locomotion, they entirely rely on diffusion and active transport along MTs to
reach the nucleus.

The cell cytoplasm is a highly crowded environment and diffusion of macromolecules depends on their size: while non interacting spherical particles
with radius up to $\approx 25nm$ are freely diffusible in the cell
cytoplasm \cite{seksek}, increasing the size above $45 nm$ reduces
considerably the motion \cite{Dauty}. Interestingly large viral
particles have developed nuclear targeting signals to be actively
transported along microtubules, which resulted in drastically decrease their arrival
time to the nucleus \cite{sodeik}. It remains a challenging question to analyze their associated viral trajectories, which consist of a succession of free or confined diffusion and/or ballistic periods \cite{Seisenberger,Arhel,brandenburg-zhuang}. The analysis of such random trajectories starts with the position ${\bf X}(t)$ at time $t$ of the gene vector, which is a stochastic process \cite{schuss,berg} and the dynamics depends on forces applied on the particle. When the motion of the gene vector is purely diffusive
such as for cationic synthetic vectors, the overdamped equation for
the velocity is simply $d{\bf X}/dt= \sqrt{2D} \,d{\bf W}/dt$, where ${\bf W}$ is the standard Brownian motion. For a viral gene vector, the motion is usually assisted by active
transport along MTs and the {stochastic} equation for the velocity becomes
\beq
\label{eq1}
\frac{d\bf{X}}{dt}={\bf b}\left( {\bf X}\right)\,+\sqrt {2D}
\,\frac{d{\bf W}}{dt},\label{equation1}
\eeq
where ${\bf b}$ is a drift that accounts for ballistic periods along
MTs. This continuous Langevin description can be used to generate
computer simulations of trajectories \cite{schuss,risken} in free
and confined environment \cite{SIAM-Holcman}, and is the basis to
derive asymptotic formulas for the probability and the mean first
passage time of the vector to a nuclear pore \cite{holcmanschuss}.
However, to reduce the complexity of large simulations, a first step
consists in studying the dependency of ${\bf b}\left( {\bf X}\right)$
as a function of the MTs organization and the viral dynamical properties
(diffusion constant $D$, affinity with microtubules and net velocity
along MTs) \cite{LH1,LH2}.

\subsubsection*{From the stochastic description of vector trajectories to the probability
and the mean arrival time to a small nuclear pore.}
The crowded cytoplasm is a risky environment for gene vectors that
can be either trapped or degraded through the cellular defense
machinery. Consequently, the cytoplasmic trafficking is rate
limiting for genes expression, and to analyze quantitatively that
step, we derived \cite{david,PRE} asymptotic expressions
for the probability $P_n$ and the mean time $\tau_n$ a single gene
vector arrives to one of the $n$ small nuclear pores. In particular,
 obtaining these expressions allows to explore the phase space of parameters and in particular 
 links global quantitative outputs measuring the success of nuclear genes delivery with the cellular geometry,
the MTs organization and the dynamical properties of gene vectors. 
To obtain these expression, we modeled the viral degradation or
immobilization by a steady state degradation rate $k({\rm {\bf
x}})$. We then introduce the survival probability density
function (SPDF) $p(\x,t)$, which is the probability to find a live (not degraded) viral particle inside a cytoplasmic volume
element $\x+d\x$ at time $t$ \cite{david}
\beq
p(\x,t) d\x= \Pr\{X(t) \in \x+ d \x, \tau^k>t ,\tau^a>t\,|\, p_i \},
 \eeq
where $\tau ^a$ is the first time for a live virus to arrive to an absorbing nuclear pore, denoted $\partial N_a $  and  $\tau ^k$ the
first time that it is degraded.  The viral initial
distribution is $p_i$. The SPDF $p(x,t)$ satisfies the Fokker-Planck
equation (FPE) \cite{schuss}
\beq
\frac{\p p(\x,t)}{\p t}&=&D\Delta p(\x,t) -
\nabla\cdot\mb{b}(\x)p(\x,t) -k(\x) p(\x,t) \quad\mbox{for}\quad\x\in\Omega\nonumber\\
p(\x,0) & = & p_i(\x)\quad\mbox{for}\quad \x\in \Omega,\label{FPE}
\eeq
which describes the time evolution of the vector probability density
function. The boundary conditions are 
\beq
p(\x,t)&=&0\quad\mbox{for}\quad\x\in\p N_a\nonumber
\label{boundary-c1}\\ \mb{J}(\x,t)\cdot\n_{\x}& =& 0\quad \x\in \p \Omega-\p N_a \label{boundary-c2},
\eeq
where the first condition describes the particle absorption at
the nuclear pores area $\p N_a$ and the second one the reflection
of the vector on the remaining boundary area of the cell $\p
\Omega-\p N_a$.  The flux density vector $\mb{J}(\x,t)$ is defined by
 \beq
\mb{J}(\x,t)=-D \nabla p(\x,t) +\mb{b}(\x)p(\x,t),\label{Ji}
 \eeq
where $\n_{\x}$ is the unit outer normal at a boundary point $\x$. The probability $P_n $ that a live virus arrives to a nuclear pore and the conditional mean time can be expressed using the SPDF as
\beq
P_n =\Pr\{\tau ^a<\tau ^k\}=\oint_{\p N_a} \int_0^\infty
\mb{J}(\x,t)\cdot\n_{\x} dS_{\x}\,dt
\eeq
and
\beq
\tau_{n}=E[\tau^a
\,|\,\tau^a<\tau^k] &=&\int_0^\infty \ds{(1-\Pr\{\tau^a
<t\,|\,\tau^a<\tau^k)}\,dt\nonumber\\&=&{P_n } {\int_0^\infty \oint_{\p N_a}
t\mb{J}(\x,t)\cdot\n_{\x} dS_{\x}\,dt}
\eeq
where $\Pr \{\tau^a<t | \tau ^a<\tau ^k\}$ is the conditional cumulative density function of the absorption time, given that the gene vector arrives alive to a nuclear pore.
For $n$ identical nuclear pores modeled as absorbing disks of radius $\epsilon$ and when the drift is a gradient potential ($\mathbf{b\left( x \right)}=-\nabla \Phi \left( \x \right)$), the leading order term of $P_n$ and $\tau_n$ in $\epsilon$ have been estimated 
\cite{david}.
\beq
P_n= \frac{\ds{e^{\ds{-\Phi_0/D}}}}
{\ds{\frac{1}{4 D
n\epsilon}\,\int_{\Omega}e^{\ds{-\Phi(\x)/D}}k(\x)d\x+  e^{\ds{-\Phi_0/D}}}} ,
\eeq
and,
\beq
\tau_n=\frac{\ds{\frac{1}{4 D
n\epsilon}\, \int_{\Omega}e^{\ds{-\Phi(\x)/D}}d\x}} 
{\ds{ \frac{1}{4 D
n\epsilon}\, \int_{\Omega}e^{\ds{-\Phi(\x)/D}}k(\x)d\x+e^{\ds{-\Phi_0/D}}}},\label{TN}
\eeq
where $\Phi_0$ is the constant value of the radial potential
$\Phi(\x)$ on the centered nucleus where the nuclear pores
are uniformly distributed.  These asymptotic results rely on the
narrow escape theory \cite{holcmanschuss}, which is a general asymptotic method to estimate the mean
first passage time of a Brownian motion confined in a domain to
escape through a small opening. 

As the number of nuclear pores $n$ remains small, the asymptotic validity of the previous expressions \ref{TN} agree in some limits with Brownian simulations \cite{PRE}. However, these formulas do not account for the possible interactions between
the small absorbing pores and indeed, $\lim_{n \rightarrow
\infty,n\epsilon^2\ll1}\langle \tau \rangle=0$. Because the
nucleus contains a large number of nuclear pores ($n=2000$
\cite{maul}) that cover a small total area (around $1\%$ of the
nuclear surface), a refined analysis has been developed to account for the nuclear
geometry \cite{lagache-many}. In addition, interaction between absorbing windows can drastically affect the
MFPT \cite{HS,Ward1}.

Using electrostatic arguments, when the number of nuclear pores
$n\gg1$ is large and cover uniformly a small surface of the nucleus
$\Sigma$, the leading order of the narrow escape time for a pure
Brownian particle is \cite{juergen}
\beq
\tau_n =\frac{|\Omega|}{D}\left(\frac{1}{C_{\Sigma}}+\frac{1}{4
n \epsilon}\right),\label{tau-st}
\eeq
where $C_{\Sigma}$ is the capacity of the nucleus (for a sphere of radius $\delta$, $C_{\Sigma}=4\pi \delta$). When we
neglected the geometrical interactions between the holes
$1/C_{\Sigma}=0$, the mean time (\ref{tau-st}) reduces to
expression (\ref{TN}) for a Brownian particle $\phi=0$ with no
killing activity $k=0$. Interestingly, both a quantitative and
qualitative information can be derived from formula (\ref{tau-st}):  the  MFPT is shorter
for a sphere covered with many small holes compared to a single large hole \cite{PRE} with same surface. Actually, for a pure
Brownian process, many absorbing small holes is equivalent to an almost
totally absorbing surface \cite{Berg-purcellBJ1977}.

Finally, in \cite{lagache-many}, {we accounted for the interactions between absorbing nuclear pores, and obtained refined estimates for the probability $P_n$ and $\tau_n$ (formula \ref{TN}), when the velocity is described by the stochastic equation (\ref{FPE})}
\beq
P_n=\frac{\ds{\frac{1}{ |\p
\Sigma|}\,\oint_{\p \Sigma}
e^{\ds{-\Phi(\x)/D}}d\x}}{\ds{\frac{1}{ |\p
\Sigma|}\, \oint_{\p \Sigma}
e^{\ds{-\Phi(\x)/D}}d\x}+\left(\frac{1}{4nD\epsilon}+\frac{1}{D C_{\Sigma}}
\right)\int_{\Omega}k(\x)e^{\ds{-\Phi(\x)/D}}d\x},\label{pfinal}
\eeq
and
\beq
\tau_n=\frac{\ds{\left(\frac{1}{4nD\epsilon}+\frac{1}{D C_{\Sigma}}\right) \int_{\Omega}e^{\ds{-\Phi(\x)/D}d\x}}}{\ds{\frac{1}{ |\p
\Sigma|}\oint_{\p \Sigma} e^{\ds{-\Phi(\x)/D}}d\x+\left(\frac{1}{4nD\epsilon}+\frac{1}{D C_{\Sigma}}\right)\int_{\Omega}k(\x)e^{\ds{-\Phi(\x)/D}}d\x}}.
\label{tfinal}
\eeq
Interestingly, for a biological cell with a spherical nucleus
(radius $\delta=5\mu m$ \cite{maul}), the  $n=2000$ circular nuclear
pores \cite{maul} (radius $\epsilon=25nm$ \cite{maul}) cover a
surface $\left(n \pi \epsilon^2\right)/\left(4 \pi \delta^2\right)
\approx 1\% $ of the total nuclear surface and
$1/\left(4nD\epsilon\right)$ is only one third of $1/C_{\Sigma}$.
Using the parameter for the Adeno-Associated-Virus, with 
an affective diffusion constant $D=1.3\mu m^2s^{-1}$ \cite{Seisenberger}
and no degradation activity ($k=0$), moving in a ball  of
radius $R=15\mu m$ (\cite{CHOovary}) containing a nucleus and a
radial potential $\Phi(r)=0.2 r \mu^2 ms^{-1}$  \cite{PRE}, using
formula \ref{tfinal},  we found that the time to a nuclear pore is  $\tau_n \approx 1min.$, which
is three time longer than the estimation using formula (\ref{TN}),
when the effect of the nucleus is not taken into
account.

\section{Conclusion}
Delivering  a plasmid DNA in a cell is still a
challenging task and a daunting hurdle of modern drug delivery
methods. Being able to quantify cytoplasmic trafficking cannot only
be used to optimize gene delivery, but also to design specific drug strategy against viral infection.

In the first part of the review, we summarized some recent progresses about quantifying
the endosomal escape of synthetic polycations vectors, which defines
a limiting step in gene delivery. It remains an open question to
understand the precise mechanisms by which a plasmid is escaping and
how the endosomal membrane is disrupted. We believe that any
understanding can lead to a rational method to optimize the
construction of polycations. In the second part, we presented some
general modeling approaches to quantify cytoplasmic trafficking of enveloped viruses,
such as influenza.  It remains to find strategies to design these new synthetic gene
vectors and for that goal, viruses appear as optimal models.  Hybrid
vectors made of a mixture of viral glycoproteins and polycations
seem to be a promising direction and physical modeling and numerical simulations are certainly 
seducing tools to optimize the construction of such nano-scale
carriers.


\subsubsection*{Acknowledgment:} D. H. research is supported by an
ERC-Starting Grant.


\end{document}